\documentclass[11pt]{article}

\usepackage{amsfonts}
\usepackage{amsmath}
\usepackage{graphicx}
\usepackage{subfigure}
\usepackage{color}
\usepackage{vmargin}
\setpapersize{USletter}
\setmarginsrb{1in}{1.5in}{1in}{0.5in}{0in}{0in}{11pt}{0.4in} % left, top, right, bottom, header, head sep, footer, foot skip
\newlength{\defbaselineskip}
\begin{document}

\title{Approximate Computation and Implicit Regularization \\ for Very 
Large-scale Data Analysis%
\footnote{To appear in the Proceedings of the 2012 ACM Symposium on 
Principles of Database Systems (PODS 2012).
}
}

\author{
Michael~W.~Mahoney%
\thanks{
Department of Mathematics, 
Stanford University, 
Stanford, CA 94305. 
Email: mmahoney@cs.stanford.edu
}
}

\date{}
\maketitle

\begin{abstract}
\noindent
Database theory and database practice are typically the domain of computer 
scientists who adopt what may be termed an algorithmic perspective on their 
data.
This perspective is very different than the more statistical perspective 
adopted by statisticians, scientific computers, machine learners, and other 
who work on what may be broadly termed statistical data analysis.
In this article, I will address fundamental aspects of this 
algorithmic-statistical disconnect, with an eye to bridging the gap between 
these two very different approaches.
A concept that lies at the heart of this disconnect is that of statistical 
regularization, a notion that has to do with how robust is the output of an 
algorithm to the noise properties of the input data.
Although it is nearly completely absent from computer science, which 
historically has taken the input data as given and modeled algorithms 
discretely, regularization in one form or another is central to nearly every 
application domain that applies algorithms to noisy data. 
By using several case studies,
I will illustrate, both theoretically and empirically, the nonobvious fact 
that approximate computation, in and of itself, can implicitly lead to 
statistical regularization.
This and other recent work suggests that, by exploiting in a more principled 
way the statistical properties implicit in worst-case algorithms, one can 
in many cases satisfy the bicriteria of having algorithms that are scalable 
to very large-scale databases and that also have good inferential or 
predictive~properties.
\end{abstract}

\section{Introduction}

Several years ago, I had the opportunity to give in several venues a 
keynote talk and to write an associated overview article on the general 
topic of ``Algorithmic and Statistical Perspectives on Large-Scale Data 
Analysis''~\cite{algstat10_CHAPTER}.
By the \emph{algorithmic perspective}, I meant roughly the approach that 
someone trained in computer science might adopt;%
\footnote{From this perspective, primary concerns include database issues, 
algorithmic questions such as models of data access, and the worst-case 
running time of algorithms for a given objective function; but there can be
a lack of appreciation, and thus associated cavalierness, when it comes to 
understanding how the data can be messy and noisy and poorly-structured in 
ways that adversely affect how confident one can be in the conclusions that 
one draws about the world as a result of the output of one's fast 
algorithms.}
and by the \emph{statistical perspective}, I meant roughly the approach that 
someone trained in statistics, or in some area such as scientific computing 
where strong domain-specific assumptions about the data are routinely made, 
might adopt.%
\footnote{From this perspective, primary concerns include questions such as 
how well the objective functions being considered conform to the phenomenon 
under study, how best to model the noise properties in the data, and whether 
one can make reliable predictions about the world from the data at hand; but 
there tends to be very little interest in understanding either computation 
\emph{per se} or the downstream effects that constraints on computation can 
have on the reliability of statistical inference.}
My main thesis was twofold.
First, motivated by problems drawn from a wide range of application domains 
that share the common feature that they generate very large quantities of 
data, we are being forced to engineer a union between these two extremely 
different perspectives or worldviews
on what the data are and what are interesting or fruitful ways to 
view the data.
Second, rather than \emph{first} making statistical modeling decisions, 
independent of algorithmic considerations, and \emph{then} applying a 
computational procedure as a black box---which is quite typical in 
small-scale and medium-scale applications and which is more natural if one
adopts one perspective or the other---in many large-scale applications 
it will be more fruitful to understand and exploit what may be termed the 
statistical properties \emph{implicit} in worst-case algorithms.
I illustrated these claims with two 
examples from
genetic and Internet 
applications; and I noted that this approach of more closely coupling the 
computational procedures used with a statistical understanding of the data 
seems particularly appropriate more generally for very large-scale data 
analysis problems. 

Here, I would like to revisit these questions, with an emphasis on 
describing in more detail particularly fruitful 
directions to 
consider in order to ``bridge the gap'' between the theory and practice of 
Modern Massive Data Set (MMDS) analysis.
On the one hand, very large-scale data are typically stored in some sort of 
database, either a variant of a traditional relational database or a 
filesystem associated with a supercomputer or a distributed cluster of 
relatively-inexpensive commodity machines.  
On the other hand, it is often noted that, in large part because they are 
typically generated in automated and thus relatively-unstructured ways, data 
are becoming increasingly ubiquitous and cheap; and also that the scarce 
resource complementary to large-scale data is the ability of the analyst to 
understand, analyze, and extract insight from those data.
As anyone who has ``rolled up the sleeves'' and worked with real data can
attest, real data are messy and noisy and poorly-structured in ways that can 
be hard to imagine before (and even sometimes after) one sees them. 
Indeed, there is often quite a bit of very practical ``heavy lifting,'' 
\emph{e.g.}, cleaning and preparing the data, to be done before starting to 
work on the ``real'' problem---to such an extent that many would say that 
big data or massive data applications are basically those for which the 
preliminary heavy lifting \emph{is} the main problem. 
This clearly places a premium on algorithmic methods that permit the analyst 
to ``play with'' the data and to work with the data interactively, as initial 
ideas are being tested and statistical hypotheses are being formed.
Unfortunately, this is not the sort of thing that is easy to do with 
traditional databases.

To address these issues, I will discuss a notion that lies at the heart of
the disconnect between the algorithmic perspective and the statistical 
perspective on data and data analysis.
This notion, often called \emph{regularization} or \emph{statistical 
regularization}, is a traditional and very intuitive idea.
Described in more detail in Section~\ref{sxn:thoughts:regularization}, 
regularization basically has to do with how robust is the output of an 
algorithm to the noise properties of the input data.
It is usually formulated as a tradeoff between ``solution quality'' (as 
measured, \emph{e.g.}, by the value of the objective function being 
optimized) and ``solution niceness'' (as measured, \emph{e.g.}, by a vector 
space norm constraint, a smoothness condition, or some other related measure 
of interest to a downstream analyst).
For this reason, when applied to noisy data, regularized objectives and 
regularized algorithms can lead to output that is ``better'' for downstream 
applications, \emph{e.g.}, for clustering or classification or other things
of interest to the domain scientist, than is the output of the corresponding 
unregularized algorithms.
Thus, although it is nearly completely absent from computer science, which 
historically has taken the input data as given and modeled algorithms 
discretely, regularization in one form or another is central to nearly every 
application domain that applies algorithms to noisy data.% 
\footnote{Clearly, there will be a problem if the output of a computer 
scientist's algorithm is manifestly meaningless in terms of the motivating 
application or if the statistician's objective function takes the age of the 
universe to optimize.
The point is that, depending on one's perspective, data are treated as a 
black box with respect to the algorithm, or vice versa; and this leads one
to formulate problems in very different ways.
From an algorithmic perspective, questions about the reliability and 
robustness of the output to noise in the input are very much secondary; and 
from a statistical perspective, the same is true regarding the details of 
the computation and the consequences of resource constraints on the 
computation.}

I will also discuss how, by adopting a very non-traditional perspective on 
approximation algorithms (or, equivalently, a non-traditional perspective on 
statistical regularization), one can in many cases satisfy the bicriteria of 
having algorithms that are scalable to very large data sets and that also 
have good statistical or inferential or predictive properties.
Basically, the non-traditional perspective is that approximate 
computation---either in the sense of approximation algorithms in 
theoretical computer science or in the sense of heuristic design decisions
(such as binning, pruning, and early stopping) that practitioners must make 
in order to implement their algorithms in real systems---often 
\emph{implicitly} leads to some sort of regularization.
That is, approximate computation, \emph{in and of itself}, can implicitly 
lead to statistical regularization.
This is very different than the usual perspective in approximation 
algorithms, where one is interested in solving a given problem, but since 
the problem is intractable one ``settles for'' the output of an 
approximation.
In particular, this means that,
depending on the details of the situation, approximate computation can 
lead to algorithms that are both faster \emph{and} better than are 
algorithms that solve the same problem exactly.

While particular examples of this phenomenon are well-known, typically 
heuristically and amongst practitioners, in my experience the general 
observation is quite surprising to both practitioners and theorists of both 
the algorithmic perspective and the statistical perspective on~data.
Thus, I will use three ``case studies'' from recent MMDS analysis to 
illustrate this phenomenon of \emph{implicit regularization via approximate 
computation} in three somewhat different ways.
The first
involves computing an approximation to the leading nontrivial eigenvector 
of the Laplacian matrix of a graph;
the second
involves computing, with two very different approximation algorithms, an 
approximate solution to a popular version of the graph partitioning problem; 
and the third
involves computing an approximation to a locally-biased version of this 
graph partitioning problem.
In each case, we will see that approximation algorithms that are run in 
practice implicitly compute smoother or more regular answers than do 
algorithms that solve the same problems exactly. 

Characterizing and exploiting the implicit regularization 
properties underlying approximation algorithms for 
large-scale data 
analysis problems is not the sort of analysis that is currently performed 
if one adopts a purely algorithmic perspective or a purely statistical 
perspective on the data.
It is, however, clearly of interest in many MMDS applications, where 
anything but scalable algorithms is out of the question, and where ignoring 
the noise properties of the data will likely lead to meaningless output.
As such, it represents a challenging interdisciplinary research front, both 
for theoretical computer science---and for database theory in 
particular---as well as for theorists and practitioners of statistical data 
analysis more generally.

\section{Some general observations \ldots}
\label{sxn:thoughts}

Before proceeding further, I would like to present in this section some 
general thoughts.
Most of these observations will be ``obvious'' to at least some readers, 
depending on their background or perspective, and most are an oversimplified 
version of a much richer story.
Nevertheless, putting them together and looking at the ``forest'' instead 
of the ``trees'' should help to set the stage for the subsequent discussion.

\subsection{\ldots\hspace{.5mm} on models of data}
\label{sxn:thoughts:models}

It helps to remember that data are whatever data are---%\emph{e.g.}, 
records of banking and other financial transactions, hyperspectral medical 
and astronomical images, measurements of electromagnetic signals in remote 
sensing applications, DNA microarray and single-nucleotide polymorphism 
measurements, term-document data from the Web, query and click logs at a 
search engine, interaction properties of users in social and information 
networks, corpora of images, sounds, videos, etc.
To do something useful with the data, one must first model them 
(either explicitly or implicitly\footnote{By implicitly, I mean that, while 
computations always return answers (yes, modulo issues associated with the 
Halting Problem, infinite loops, etc.), in many cases one can say that a 
given computation is the ``right'' thing to do for a certain class of data.  
For example, performing matrix-based computations with $\ell_2$-based 
objectives often has an interpretation in terms of underlying Gaussian 
processes.  Thus, performing that computation in some sense implicitly 
amounts to assuming that that is what the data ``look like.''}) 
in some way.
At root, a \emph{data model} is a mathematical structure such that---given 
hardware, communication, input-output, data-generation, sparsity, noise, 
etc. considerations---one can perform computations of interest to yield 
useful insight on the data and processes generating the data. 
As such, choosing an appropriate data model has algorithmic, statistical, 
and implementational aspects that are typically intertwined in complicated 
ways.
Two criteria to keep in mind in choosing a data model are the following.
\begin{itemize}
\item
First, on the \emph{data acquisition or data generation side}, one would like 
a structure that is ``close enough'' to the data, \emph{e.g.}, to the processes 
generating the data or to the noise properties of the data or to natural 
operations on the data or to the way the data are stored or accessed, that 
modeling the data with that structure does not do too much ``damage'' to 
the~data.
\item
Second, on the \emph{downstream or analysis side}, one would like a 
structure that is at a ``sweet spot'' between descriptive flexibility and 
algorithmic tractability.
That is, it should be flexible enough that it can describe a range of types 
of data, but it should not be so flexible that it can do ``anything,'' in 
which case computations of interest will likely be intractable and inference 
will be problematic. 
\end{itemize}

\noindent
Depending on the data and applications to be considered, the data may be 
modeled in one or more of several ways.
\begin{itemize}
\item
\emph{Flat tables and the relational model.}
Particularly common in database theory and practice, this model views the 
data as one or more two-dimensional arrays of data elements.  
All members of a given column are assumed to be similar values; all members 
of a given row are assumed to be related to one another; and different 
arrays can be related to one another in terms of predicate logic and set 
theory, which allows one to query, \emph{e.g.}, with SQL or a variant, the 
data.
\item
\emph{Graphs, including special cases like trees and expanders.}
This model is particularly common in computer science theory and practice; 
but it is also used in statistics and machine learning, as well as in 
scientific computation, where it is often viewed as a discretization of an
underlying continuous problem.
A graph $G=(V,E)$ consists of a set of vertices $V$, that can represent some 
sort of ``entities,'' and a set of edges $E$, that can be used to represent 
pairwise ``interactions'' 
between two entities.
There is a natural geodesic distance between pairs of vertices, which 
permits the use of ideas from metric space theory to develop algorithms; and 
from this perspective natural operations include breadth-first search and 
depth-first search.
Alternatively, in spectral graph theory, eigenvectors and eigenvalues of 
matrices associated with the graph are of interest;
and from this perspective, one can consider resistance-based or 
diffusion-based notions of distance between pairs of vertices.
\item
\emph{Matrices, including special cases like symmetric positive semidefinite 
matrices.} 
An $m \times n$ real-valued matrix $A$ provides a natural structure for 
encoding information about $m$ objects, each of which is described by $n$ 
features; or, if $m=n$, information about the correlations between all $m$
objects. 
As such, this model is ubiquitous in areas of applied mathematics such as 
scientific computing, statistics, and machine learning, and it is of 
increasing interest in theoretical computer science.
Rather than viewing a matrix simply as an $m \times n$ array of numbers, one 
should think of it as representing a linear transformation between two 
Euclidean spaces, $\mathbb{R}^{n}$ and $\mathbb{R}^{m}$; 
and thus vector space concepts like dot products, orthogonal matrices, 
eigenvectors, and eigenvalues are natural.
In particular, matrices have a very different semantics than tables in the 
relational model, and Euclidean spaces are much more structured objects 
than arbitrary metric spaces. 
\end{itemize}
Of course, there are other ways to model data---\emph{e.g.}, DNA sequences 
are often fruitfully modeled by strings---but matrices and graphs are most
relevant to our discussion~below.

Database researchers are probably most familiar with the basic flat table 
and the relational model and its various extensions; and there are many 
well-known advantages to working with them.
As a general rule, these models and their associated logical operations 
provide a powerful way to process the data at hand; but they are much less 
well-suited for understanding and dealing with imprecision and the noise 
properties in that data.
(See~\cite{claremont08,madskills09} and references therein.)
For example, historically, the focus in database theory and practice has 
been on business applications, \emph{e.g.}, automated banking, corporate 
record keeping, airline reservation systems, etc., where requirements such 
as performance, correctness, maintainability, and reliability (as opposed to 
prediction or inference) are crucial.

The reason for considering more sophisticated or richer data models is 
that much of the ever-increasing volume of data that is currently being 
generated is either relatively-unstructured or large and internally 
complex in its original form; and many of these 
noisy unstructured data are better-described by (typically sparse and 
poorly-structured) graphs or matrices than as dense flat tables.  
While this may be obvious to some, 
the graphs and matrices that arise in MMDS applications are very different
than those arising in classical graph theory and traditional numerical 
linear algebra; and thus
modeling large-scale% 
\footnote{Clearly, large or big or massive means different things to 
different people in different applications.
Perhaps the most intuitive description is that one can call the size of a 
data set: 
\emph{small} if one can look at the data, fairly obviously see a good 
solution to problems of interest, and find that solution fairly easily 
with almost any ``reasonable'' algorithmic tool;
\emph{medium} if the data fit in the RAM on a reasonably-priced laptop or 
desktop machine and if one can run computations of interest on the data in 
a reasonable length of time; %%  without parallel or distributed computation;
and
\emph{large} if the data doesn't fit in RAM or if one can't 
relatively-easily run computations of interest in a reasonable length of 
time. 
The point is that, as one goes from medium-sized data to large-scale data 
sets, the main issue is that one doesn't have random access to the data, and 
so details of communication, memory access, etc., become paramount~concerns.}
data by graphs and 
matrices poses very substantial challenges, given the way that databases (in 
computer science) have historically been constructed, the way that 
supercomputers (in scientific computing) have historically been designed, 
the tradeoffs that are typically made between faster CPU time and better IO 
and network communication, etc.

\subsection{\ldots\hspace{.5mm} on the relationship between algorithms and data}
\label{sxn:thoughts:relationship}

Before the advent of the digital computer, the natural sciences (and to a 
lesser extent areas such as social and economic sciences) provided a rich 
source of problems; and statistical methods were developed in order to solve 
those problems.
Although these statistical methods typically involved computing something, 
there was less interest in questions about the nature of computation 
\emph{per se}.
That is, although computation was often crucial, it was in some sense 
secondary to the motivating downstream application.
Indeed, an important notion was (and still is) that of a \emph{well-posed 
problem}---roughly, a problem is well-posed if: a solution exists; that 
solution is unique; and that solution depends continuously on the input data 
in some reasonable topology.
Especially in numerical applications, such problems are sometimes 
called \emph{well-conditioned problems}.%
\footnote{In this case, the \emph{condition number} of a problem, which 
measures the worst-case amount that the solution to the problem changes when 
there is a small change in the input data, is small for well-conditioned 
problems.}
From this perspective, it simply doesn't make much sense to consider 
algorithms for problems that are not well-posed---after all, any possible 
algorithm for such an ill-posed problem will return answers that are not be 
meaningful in terms of the domain from which the input data are drawn.

With the advent of the digital computer, there occurred a split in the 
yet-to-be-formed field of computer science.
The split was loosely based on the application domain (scientific computing 
and numerical computation versus business and consumer applications), but 
relatedly based on the type of tools used (continuous mathematics like 
matrix analysis and probability versus discrete mathematics like 
combinatorics and logic); and it led to two very different perspectives 
(basically the statistical and algorithmic perspectives) on the relationship 
between algorithms and data.

On the one hand, for many numerical problems that arose in applications of 
continuous mathematics, a two-step approach was used.
It turned out that, even when working with a given well-conditioned problem,%
\footnote{Thus, the first step is to make sure the problem being considered 
is well-posed.  Replacing an ill-posed problem with a related well-posed 
problem is common and is, as I will describe in 
Section~\ref{sxn:thoughts:regularization}, a form of regularization.}
certain algorithms that solved that problem ``exactly'' in some idealized 
sense performed very poorly in the presence of ``noise'' introduced by the 
peculiarities of roundoff and truncation errors.
Roundoff errors have to do with representing real numbers with only 
finitely-many bits; and truncation errors arise since only a finite number 
of iterations of an iterative algorithm can actually be performed.
The latter are important even in ``exact arithmetic,'' since most problems 
of continuous mathematics cannot even in principle be solved by a finite 
sequence of elementary operations; and thus, from this perspective, 
fast algorithms are those that converge quickly to approximate answers that 
are accurate to, \emph{e.g}, $2$ or $10$ or $100$ digits of~precision.

This led to the notion of the \emph{numerical stability} of an algorithm.
Let us view a numerical algorithm as a function $f$ attempting to map the 
input data $x$ to the ``true'' solution $y$; but due to roundoff and 
truncation errors, the output of the algorithm is actually some other $y^*$.
In this case, the \emph{forward error} of the algorithm is 
$\Delta y = y^* - y$; and the \emph{backward error} of the algorithm is the 
smallest $\Delta x$ such that $f(x + \Delta x) = y^*$.
Thus, the forward error tells us the difference between the exact or true 
answer and what was output by the algorithm; and the backward error tells 
us what input data the algorithm we ran actually solved exactly.
Moreover, the forward error and backward error for an algorithm 
are related by the condition number of the problem---the magnitude of the 
forward error is bounded above by the 
condition number multiplied by the 
magnitude of the backward error.%
\footnote{My apologies to those readers who went into computer science, and 
into database theory in particular, to avoid these sorts of numerical issues, 
but these distinctions 
really do matter
for what I will be describing below.} 
In general, a backward stable algorithm can be expected to provide an 
accurate solution to a well-conditioned problem; and much of the work in 
numerical analysis, continuous optimization, and scientific computing can 
be seen as an attempt to develop algorithms for well-posed problems that 
have better stability properties than the ``obvious'' unstable~algorithm.

On the other hand, it turned out to be much easier to study computation 
\emph{per se} in discrete settings (see~\cite{Sma90,BCSS96} for a partial 
history), and in this case a simpler but coarser one-step approach prevailed.
First, several seemingly-different approaches (recursion theory, the 
$\lambda$-calculus, and Turing machines) defined the same class of functions.
This led to the belief that the concept of computability is formally captured 
in a qualitative and robust way by these three equivalent processes, 
independent of the input data; and this highlighted the central role of 
logic in this approach to the study of computation.
Then, it turned out that the class of computable functions has a rich 
structure---while many problems are solvable by algorithms that run in 
low-degree polynomial time, some problems seemed not to be solvable by 
anything short of a trivial brute force algorithm.
This led to the notion of the complexity classes P and NP, the concepts of 
NP-hardness and NP-completeness, etc., the success of which led to the 
belief that the these classes formally capture in a qualitative and robust 
way the concept of computational tractability and intractability, 
independent of any posedness questions or any assumptions on the input data.

Then, it turned out that many problems of practical interest are 
intractable---either in the sense of being NP-hard or NP-complete or, of more
recent interest, in the sense of requiring $O(n^2)$ or $O(n^3)$ time when 
only $O(n)$ or $O(n \log n)$ time is available.
In these cases, computing some sort of approximation is typically of interest.
The modern theory of approximation algorithms, as formulated in theoretical
computer science, provides forward error bounds for such problems for 
``worst-case'' input.
These bounds are worst-case in two senses: first, they hold uniformly for all 
possible input; and second, they are typically stated in terms of a 
relatively-simple complexity measure such as problem size, independent of any
other structural parameter of the input data.%
\footnote{The reason for not parameterizing running time and approximation 
quality in terms of structural parameters 
is that one can 
encode all sorts of pathological things in combinatorial 
parameters, thereby obtaining trivial results.} 
While there are several ways to prove worst-case bounds for approximation 
algorithms, a common procedure is to take advantage of 
relaxations---\emph{e.g.}, solve a relaxed linear program, rather than an 
integer program formulation of the combinatorial 
problem~\cite{Vazirani01,HLW06_expanders}.
This essentially involves ``filtering'' the input data through some other 
``nicer,'' often convex, metric or geometric space.
Embedding theorems and duality then bound how much the input data are
distorted by this filtering and provide worst-case quality-of-approximation 
guarantees~\cite{Vazirani01,HLW06_expanders}.

\subsection{\ldots\hspace{.5mm} on explicit and implicit regularization}
\label{sxn:thoughts:regularization}

The term \emph{regularization} refers to a general class of 
methods~\cite{Neu98,CH02,BL06} to ensure that the output of an algorithm is 
meaningful in some sense---\emph{e.g.}, 
to the domain scientist who is interested in using that output for some 
downstream application of interest in the domain from which the data are 
drawn; to someone who wants to avoid ``pathological'' solutions; or to a 
machine learner interested in prediction accuracy or some other form of 
inference. 
It typically manifests itself by requiring that the output of an algorithm 
is not overly sensitive to the noise properties of the input data; and, as a 
general rule, it provides a tradeoff 
between the quality and the niceness of the solution.

Regularization arose in integral equation theory where there 
was interest in providing meaningful solutions to ill-posed 
problems~\cite{TikhonovArsenin77}.
A common approach was to assume a smoothness condition on the solution or to 
require that the solution satisfy a vector space norm constraint.
This approach is followed much more generally in modern statistical data 
analysis~\cite{hast-tibs-fried}, where the 
posedness question
has to do with how meaningful it is to run a given algorithm, 
given the noise properties of the data, if the goal is to predict well on 
unseen data.
One typically considers a loss function $f(x)$ that specifies an ``empirical 
penalty'' depending on both the data and a parameter vector $x$; and a 
regularization function $g(x)$ 
that provides ``geometric capacity control'' on the vector~$x$.
Then, rather than minimizing $f(x)$ exactly, one exactly solves an 
optimization problem of the~form:
\begin{equation}
\label{eqn:reg-gen}
\hat{x} = \mbox{argmin}_x f(x) + \lambda g(x)   ,
\end{equation}
where the parameter $\lambda$ intermediates between solution quality and 
solution niceness.
Implementing regularization explicitly in this manner leads to a natural 
interpretation in terms of a trade-off between optimizing the objective and 
avoiding over-fitting the data; and it can often be given a Bayesian
statistical interpretation.%
\footnote{Roughly, such an interpretation says that if the data are 
generated according to a particular noise model, then $g(\cdot)$ encodes 
``prior assumptions'' about the input data, and regularizing with this 
$g(\cdot)$ is the ``right'' thing to do~\cite{hast-tibs-fried}.}
By optimizing exactly a combination of two functions, 
though,
regularizing in this way often leads to optimization problems that are harder 
(think of $\ell_1$-regularized $\ell_2$-regression) or at least no easier 
(think of $\ell_2$-regularized $\ell_2$-regression) than the original 
problem, a situation that is clearly unacceptable in many MMDS~applications.

On the other hand, regularization is often observed as a side-effect or 
by-product of other design decisions.%
\footnote{See~\cite{Neu98,CH02,BL06,hast-tibs-fried,MO11-implementing} and 
references therein for more details on these examples.}
For example, 
``binning'' is often used to aggregate the data into bins, upon which
computations are performed;   %%%~\cite{SY05};
``pruning'' is often used to remove sections of a decision tree that 
provide little classification power;   %%%~\cite{XXX-XXX};
taking measures to improve numerical properties can also penalize large 
weights (in the solution vector) that exploit correlations beyond the
level of precision in the data generation process; and    %%%~\cite{SR03}; and
``adding noise'' to the input data before running a training algorithm can 
be equivalent to Tikhonov regularization.   %%%~\cite{Bis95}.
More generally, it is well-known amongst practitioners that certain 
heuristic approximations that are used to speed up computations can 
also have the empirical side-effect of performing smoothing or regularization.  
For example,
working with a truncated singular value decomposition in latent factor 
models can lead to better precision and recall;   %%%~\cite{XXX-XXX};
``truncating'' to zero small entries or ``shrinking'' all entries of a 
solution vector is common in iterative algorithms; and   %%%~\cite{XXX-XXX}; and
``early stopping'' is often used when a learning model such as a neural 
network is trained by an iterative gradient descent algorithm.  %%%~\cite{ZY05}.

Note that in addition to its use in making ill-posed problems 
well-posed---a distinction that is not of interest in the study of 
computation \emph{per se},
where a sharp dividing line is drawn between algorithms and input data, 
thereby effectively assuming away the posedness problem---the use of 
regularization blurs the rigid lines between algorithms and input data in 
other ways.%
\footnote{In my experience, researchers who adopt the algorithmic 
perspective are most comfortable when given a well-defined problem, in which 
case they develop algorithms for that problem and ask how those algorithms 
behave on the worst-case input they can imagine.
Researchers who adopt the statistical perspective will 
note that formulating the problem is typically the hard part; and that, if 
a problem is meaningful and well-posed, then often several related 
formulations will behave similarly for downstream applications, in a manner 
quite unrelated to their worst-case~behavior.}
For example, in addition to simply modifying the objective function to be 
optimized, regularization can involve adding to it various smoothness 
constraints---some of which involve modifying the objective and then 
calling a black box algorithm, but some of which are more simply enforced by 
modifying the steps of the original algorithm.
Similarly, binning and pruning can be viewed as preprocessing the 
data, but they can also be implemented inside the algorithm; and
adding noise to the input before running a training algorithm is clearly a 
form of preprocessing, but empirically similar regularization effects are 
observed when randomization is included inside the algorithm, \emph{e.g.}, 
as with randomized algorithms for matrix problems such as low-rank matrix 
approximation and least-squares approximation~\cite{Mah-mat-rev_BOOK}.
Finally, truncating small entries of a solution vector to zero in an 
iterative algorithm and performing early stopping in an iterative algorithm 
are clearly heuristic approximations that lead an algorithm to compute some 
sort of approximation to the solution that would have been computed had the
truncation and early stopping not been performed.

\section{Three examples of implicit regularization}
\label{sxn:comp-reg}

In this section, I will discuss three 
case studies
that illustrate the
phenomenon of implicit regularization via approximate computation in three
somewhat different ways.
For each of these problems, there exists strong underlying theory; and there
exists the practice, which typically involves approximating the exact 
solution in one way or another.
Our goal will be to understand the differences between the theory and the
practice in light of the discussion from Section~\ref{sxn:thoughts}.
In particular, rather than being interested in the output of the 
approximation procedure insofar as it provides an approximation to the exact 
answer, we will be more interested in what the approximation algorithm 
actually computes, whether that approximation can be viewed as a smoother or 
more regular version of the exact answer, and how much more generally in 
database theory and practice similar ideas can be applied.

\subsection{Computing the leading nontrivial eigenvector of a Laplacian matrix}
\label{sxn:eigenvector}

The problem of computing eigenvectors of the Laplacian matrix of a graph 
arises in many data analysis applications, including (literally) for 
Web-scale data matrices.
For example, the leading nontrivial eigenvector, \emph{i.e.}, the 
eigenvector, $v_2$, associated with the smallest non-zero eigenvalue,
 $\lambda_2$, is often of interest:
it defines the slowest mixing direction for the natural random walk on the 
graph, and thus it can be used in applications such as viral marketing,
rumor spreading, and graph partitioning;
it can be used for classification and other common machine learning tasks; 
and variants of it provide ``importance,'' ``betweenness,'' and ``ranking'' 
measures for the nodes in a graph.
Moreover, computing this eigenvector is a problem for which there exists a 
very clean theoretical characterization of how approximate computation can 
implicitly lead to statistical regularization.

Let $A$ be the adjacency matrix of a connected, weighted, undirected
graph $G=(V,E)$, and let $D$ be its diagonal degree matrix. 
That is, $A_{ij}$ is the weight of the edge between the $i^{th}$ node and 
the $j^{th}$ node, and $D_{ii}=\sum_{j:(ij)\in E} A_{ij}$.
The \emph{combinatorial Laplacian} of $G$ is the matrix $L = D-A$.
Although this matrix is defined for any graph, it has strong connections 
with the Laplace-Beltrami operator on Riemannian manifolds in Euclidean 
spaces.
Indeed, if the graph is a discretization of the manifold, then the former 
approaches the latter, under appropriate sampling and regularity 
assumptions.
In addition, the \emph{normalized Laplacian} of~$G$ is 
$\mathcal{L} = D^{-1/2}LD^{-1/2} =  I - D^{-1/2}AD^{-1/2}$.
This degree-weighted Laplacian is more appropriate for graphs with 
significant degree variability, in large part due to its connection with 
random walks and other diffusion-based processes.
For an $n$ node graph, $\mathcal{L}$ is an $n \times n$ positive 
semidefinite matrix, \emph{i.e.}, all its eigenvalues 
$\lambda_1 \le \lambda_2 \le \cdots \le \lambda_n$
are nonnegative, 
and for a connected graph, $\lambda_1 = 0$ and $\lambda_2 >0$.
In this case, the degree-weighted all-ones vector, \emph{i.e.}, the 
vector whose $i^{th}$ element equals (up to a possible normalization) 
$D_{ii}$ and which is often denoted $v_1$, is an eigenvector of 
$\mathcal{L}$ with eigenvalue zero, \emph{i.e.}, $\mathcal{L} v_1 = 0 v_1$.
For this reason, $v_1$ is often called trivial eigenvector of $\mathcal{L}$, 
and it is the next eigenvector that is of interest.  
This leading nontrivial eigenvector, $v_2$, is that vector that optimizes 
the Rayleigh quotient, defined to be $x^T\mathcal{L}x$ for a unit-length 
vector $x$, over all vectors perpendicular to the trivial eigenvector.%
\footnote{Eigenvectors of $\mathcal{L}$ can be related to generalized 
eigenvectors of $L$: if $\mathcal{L}x = \lambda x$, then $Ly = \lambda D y$, 
where $y=D^{-1/2}x$.}

In most applications where this leading nontrivial eigenvector is of 
interest, other vectors can also be used.
For example, if $\lambda_2$ is not unique then $v_2$ is not uniquely-defined 
and thus the problem of computing it is not even well-posed; 
if $\lambda_3$ is very close to $\lambda_2$, then any vector in the subspace 
spanned by $v_2$ and $v_3$ is nearly as good (in the sense of forward 
error or objective 
function value) as $v_2$; and, more generally, \emph{any} vector can be used 
with a quality-of-approximation loss that depends on how far it's Rayleigh 
quotient is from the Rayleigh quotient of $v_2$.
For most small-scale and medium-scale applications, this vector $v_2$ is 
computed ``exactly'' by calling a black-box solver.%
\footnote{To the extent, as described in 
Section~\ref{sxn:thoughts:relationship}, that any numerical computation can
be performed ``exactly.''}
It could, however, be approximated with an iterative method such as the 
Power Method%
\footnote{The Power Method takes as input any $n \times n$ symmetric matrix 
$A$ and returns as output a number $\lambda$ 
and a vector 
$v$ 
such that $Av = \lambda v$.
It starts with an initial 
vector, 
$\nu_0$, and it iteratively 
computes $\nu_{t+1} = A \nu_t /||A\nu_t||_2$.
Under weak assumptions, it converges to $v_{max}$, the dominant eigenvector 
of $A$.
The reason is clear: if we expand $\nu_0 = \sum_{i=1}^{n} \gamma_i v_i$ 
in the basis provided by the eigenfunctions $\{v_i\}_{i=1}^{n}$ of $A$, then 
$\nu_t = \sum_{i=1}^{n} \gamma_i^t v_i \rightarrow v_{max}$.
Vanilla versions of the Power Method can easily be improved (at least when 
the entire matrix $A$ is available in RAM) to obtain better stability and 
convergence properties; but these more sophisticated eigenvalue algorithms 
can often be viewed as variations of it.  
For instance, Lanczos algorithms look at a subspace of vectors generated 
during the iteration.}
or by running a random walk-based or diffusion-based procedure; 
and in many larger-scale applications this is~preferable.

Perhaps the most well-known example of this is the computation of the 
so-called PageRank of the Web graph~\cite{PBMW99}.
As an example of a spectral ranking method~\cite{Vig09_TR}, PageRank provides
a ranking or measure of importance for a Web page; and the Power Method has 
been used extensively to perform very large-scale PageRank 
computations~\cite{berkhin05_pagerank}.
Although it was initially surprising to many,
the 
Power Method has several well-known advantages for 
such Web-scale computations: 
it can be implemented with simple matrix-vector multiplications, thus not 
damaging the sparsity of the (adjacency or Laplacian) matrix; 
those matrix-vector multiplications are strongly parallel, permitting one to 
take advantage of parallel and distributed environments (indeed, MapReduce 
was originally developed to perform related Web-scale 
computations~\cite{DG04}); 
and the algorithm is simple enough that it can be ``adjusted'' and 
``tweaked'' as necessary, based on systems considerations and other design 
constraints. 
Much more generally, other spectral ranking procedures compute vectors that 
can be used instead of the second eigenvector $v_2$ to perform ranking, 
classification, clustering, etc.~\cite{Vig09_TR}.

At root, these random walk or diffusion-based methods assign positive and/or 
negative ``charge'' (or relatedly probability mass) to the nodes, and then 
they let the distribution evolve according to dynamics derived from the 
graph structure.
Three canonical evolution dynamics are the following.
\begin{itemize}
  \item \textbf{Heat Kernel.}
    Here, the charge evolves according to the 
    %% dynamics of the 
    heat equation
    $\frac{\partial H_t}{\partial t} = - \mathcal{L} H_t$.
    That is, the vector of charges evolves as
$     H_t = \exp ( -t\mathcal{L} )  = \sum_{k=0}^{\infty} \frac{(-t)^k}{k!}\mathcal{L}^k  $,
    where $t \ge 0$ is a time parameter, times an input seed 
    distribution vector.
  \item \textbf{PageRank.}
    Here, the charge evolves by either moving to a neighbor of 
    the current node or teleporting to a random node.
    That is, the vector of charges evolves as 
    \begin{equation}
    \label{eqn:page-rank}
    R_{\gamma} = \gamma \left(I-\left(1-\gamma \right)M \right)^{-1}   ,
    \end{equation}
    where $M = AD^{-1}$ is the natural random walk transition matrix associated 
    with the graph and
    where $\gamma \in (0,1)$ is the so-called teleportation parameter,
    times an input seed vector.
  \item \textbf{Lazy Random Walk.}
    Here, the charge either stays at the current node or moves to a neighbor.
    That is, if $M$ is the natural random walk transition matrix,
    then the vector of charges evolves as some power of 
$     W_{\alpha}= \alpha I + (1-\alpha)M $,
    where $\alpha \in (0,1)$ represents the ``holding probability,'' times 
    an input seed vector.
\end{itemize}

\noindent
In each of these cases, there is an input ``seed'' distribution vector,
and there is a parameter ($t$, $\gamma$, and the number 
of steps of the Lazy Random Walk) that controls the ``aggressiveness'' of 
the dynamics and thus how quickly the diffusive process equilibrates.
In many applications, one chooses the initial seed distribution carefully% 
\footnote{In particular, if one is interested in global spectral graph 
partitioning, as in Section~\ref{sxn:partitioning}, then this seed vector 
could have randomly positive entries or could be a vector with entries drawn 
from $\{-1,+1\}$ uniformly at random; 
while if one is interested in local spectral graph 
partitioning~\cite{Spielman:2004,andersen06local,Chung07_heatkernelPNAS,MOV09_TRv3},
as in Section~\ref{sxn:local}, then this vector could be the indicator 
vector of a small ``seed set'' of nodes.}
and/or prevents the diffusive process from equilibrating to the asymptotic 
state.
(That is, if one runs any of these diffusive dynamics to a limiting 
value of the aggressiveness parameter, then under weak assumptions an exact 
answer is computed, independent of the initial seed vector; but if one 
truncates  this process early, then some sort of approximation, which in 
general depends strongly on the initial seed set, is computed.)
The justification for doing this is typically that it is too expensive or 
not possible to solve the problem exactly; that the resulting 
approximate answer
has 
good forward error bounds on it's Rayleigh quotient;
and that, for many downstream applications, the resulting vector is even 
better (typically in some sense that is not precisely described) than the 
exact answer.

To formalize this last idea
in the context of classical regularization 
theory, let's ask what these approximation procedures actually compute.
In particular, do these diffusion-based approximation methods
exactly optimize a regularized objective of the 
form of Problem~(\ref{eqn:reg-gen}), where $g(\cdot)$ is nontrivial, 
\emph{e.g.}, some well-recognizable function or at least something that is 
``little-o'' of the length of the source code, and where $f(\cdot)$ is the 
Rayleigh quotient?

To answer this question, recall that $v_2$ exactly solves the following 
optimization problem.
\begin{equation}
\begin{aligned}
  & \underset{x}{\text{minimize}}
  & & x^T\mathcal{L}x \\ 
  & \text{subject to}
  & & x^Tx = 1, \\
  & & & x^T D^{1/2} 1 = 0  .
\end{aligned}
\label{eqn:mo-unreg-vp}
\end{equation}
The solution to Problem~(\ref{eqn:mo-unreg-vp}) can also be characterized as 
the solution to the following SDP (semidefinite program).
\begin{equation}
\begin{aligned}
  & \underset{X}{\text{minimize}}
  & & \mathrm{Tr}(\mathcal{L} X) \\ 
  & \text{subject to}
  & & X \succeq 0, \\
  & & & \mathrm{Tr}(X) = 1, \\
  & & & X D^{1/2} 1 = 0,
\end{aligned}
\label{eqn:mo-unreg-sdp}
\end{equation}
where $\mathrm{Tr}(\cdot)$ stands for the matrix Trace operation.
Problem~(\ref{eqn:mo-unreg-sdp}) is a relaxation of 
Problem~(\ref{eqn:mo-unreg-vp}) from an 
optimization over unit vectors to an optimization over distributions over 
unit vectors, represented by the density matrix $X$.
These two programs are equivalent, however, in that the solution to 
Problem~(\ref{eqn:mo-unreg-sdp}), call it $X^{*}$, is a rank-one matrix, 
where the vector into which that matrix decomposes, call it $x^{*}$, is 
the solution to Problem~(\ref{eqn:mo-unreg-vp}); and vice versa.

Viewing $v_2$ as the solution to an SDP makes it easier to address the 
question of what is the objective that approximation algorithms for 
Problem~(\ref{eqn:mo-unreg-vp}) are solving exactly.
In particular,
it can be shown that these three diffusion-based dynamics 
arise as solutions to the following regularized~SDP.
\begin{equation}
\begin{aligned}
  & \underset{X}{\text{minimize}}
  & & \mathrm{Tr}(\mathcal{L} X) + \tfrac{1}{\eta} G(X) \\
  & \text{subject to}
  & & X \succeq 0, \\
  & & & \mathrm{Tr}(X) = 1, \\
  & & & X D^{1/2} 1 = 0,
\end{aligned}
\label{eqn:mo-reg-sdp}
\end{equation}
where $G(\cdot)$ is a regularization function, which is the generalized 
entropy, the log-determinant, and a certain matrix-$p$-norm, 
respectively~\cite{MO11-implementing}; and where $\eta$ is a parameter 
related to the aggressiveness of the diffusive 
process~\cite{MO11-implementing}.
Conversely, solutions to the regularized SDP of 
Problem~(\ref{eqn:mo-reg-sdp}) for appropriate values of $\eta$ can be 
computed \emph{exactly} by running one of the above three diffusion-based 
approximation algorithms.
Intuitively, $G(\cdot)$ is acting as a penalty function, in a manner 
analogous to the $\ell_2$ or $\ell_1$ penalty in Ridge regression or Lasso 
regression, respectively; and by running one of these three dynamics one is 
\emph{implicitly} making assumptions about the functional form of $G(\cdot)$.%
\footnote{For readers interested in statistical issues, I should note that 
one can give a statistical framework to provide a Bayesian interpretation 
that makes this intuition precise~\cite{PM11}.  Readers not interested in 
statistical issues should at least know that these assumptions are 
implicitly being made when one runs such an approximation algorithm.}
More formally, 
this result provides a very clean theoretical characterization 
of how each of these three approximation algorithms
for computing an approximation to the leading nontrivial 
eigenvector of a graph Laplacian can be seen as exactly optimizing a 
regularized version of the same problem.

\subsection{Graph partitioning}
\label{sxn:partitioning}

Graph partitioning refers to a family of objective functions and associated
approximation algorithms that involve cutting or partitioning the nodes of a 
graph into two sets with the goal that the cut has good quality 
(\emph{i.e.}, not much edge weight crosses the cut) as well as good balance 
(\emph{i.e.}, each of the two sets has a lot of the node weight).%
\footnote{There are several standard formalizations of this bi-criterion, 
\emph{e.g.}, the graph bisection problem, the $\beta$-balanced cut problem, 
and quotient cut formulations. 
In this article, I will be interested in conductance, which is a quotient 
cut formulation, but variants of most of what I say will hold for the other 
formulations.}
As such, it 
has been studied from 
a wide range of perspectives and in a wide range of applications.
For example, it has been studied for years in scientific 
computation (where one is interested in load balancing in parallel computing 
applications), machine learning and computer vision (where one is interested 
in segmenting images and clustering data), and theoretical computer science 
(where one is interested in it as a primitive in divide-and-conquer 
algorithms); and more recently it has been studied in the analysis of large 
social and information networks (where one is interested in finding 
``communities'' that are meaningful in a domain-specific context or in 
certifying that no such communities exist).

Given an undirected, possibly weighted, graph $G=(V,E)$, the 
\emph{conductance $\phi(S)$ of a set of nodes $S \subset V$} is: 
\begin{equation}
\phi(S) = \frac{ |E(S, \overline{S})| }{ \min\{A(S),A(\bar{S})\} } ,
\label{eqn:conductance_set}
\end{equation}
where $E(S, \overline{S})$ denotes the set of edges having one end in $S$ 
and one end in the complement $\overline{S}$; where $|\cdot|$ denotes 
cardinality (or weight); where $A(S)=\sum_{i \in S} \sum_{j \in V} A_{ij}$; 
and where $A$ is the adjacency matrix of a graph.%
\footnote{For readers more familiar with the concept of expansion, where 
the \textit{expansion $\alpha(S)$ of a set of nodes $S \subseteq V$} is
$\alpha(S) = |E(S, \overline{S})| / \min\{|S|,|\overline{S}|)\}$,
the conductance is simply a degree-weighted version of the expansion.}
In this case, the \textit{conductance of the graph $G$} is:
\begin{equation}
\phi(G) = \min_{S \subseteq V} \phi(S)  .
\label{eqn:conductance_graph}
\end{equation}

\noindent
Although exactly solving the combinatorial 
Problem~(\ref{eqn:conductance_graph}) is intractable, 
there are a wide range of heuristics and approximation 
algorithms, the respective strengths and weaknesses of which are 
well-understood in theory and/or practice, for approximately optimizing 
conductance. 
Of particular interest here are \emph{spectral methods} and 
\emph{flow-based methods}.%
\footnote{Other methods include local improvement methods, which can be used 
to clean up partitions found with other methods, and multi-resolution 
methods, which can view graphs at multiple size scales.  Both of these are 
important in practice, as vanilla versions of spectral algorithms and 
flow-based algorithms can easily be improved with them.}

Spectral algorithms compute an approximation to 
Problem~(\ref{eqn:conductance_graph}) by solving 
Problem~(\ref{eqn:mo-unreg-vp}), either exactly or approximately, and then 
performing a ``sweep cut'' over the resulting vector.
Several things are worth noting.
\begin{itemize}
\item
First, Problem~(\ref{eqn:mo-unreg-vp}) is a relaxation of 
Problem~(\ref{eqn:conductance_graph}), as can be seen by replacing the 
$x\in\{-1,1\}^{n}$ constraint in the corresponding integer program with the 
constraint $x\in\mathbb{R}^{n}$ subject to $x^Tx=1$, \emph{i.e.}, by 
satisfying the combinatorial constraint ``on average''.
\item
Second, this relaxation effectively embeds the data on the one-dimensional%%
\footnote{One can also view this as ``embedding'' a scaled version of the 
complete graph into the input graph.
This follows from the SDP formulation of Problem~(\ref{eqn:mo-unreg-sdp}); 
and this is of interest since a complete graph is like a constant-degree 
expander---namely, a metric space that is ``most unlike'' low-dimensional 
Euclidean spaces such as one-dimensional lines---in terms of its cut 
structure~\cite{LLR95_JRNL,Leighton:1999}.
This provides tighter duality results, and the reason for this connection is 
that the identity on the space perpendicular to the degree-weighted all-ones 
vector is the Laplacian matrix of a complete graph~\cite{MOV09_TRv3}.}
span of $v_2$---although, since the distortion is minimized only on average, 
there may be some pairs of points that are distorted a lot. 
\item
Third, one can prove that the resulting partition is ``quadratically good,'' 
in the sense that the cut returned by the algorithm has conductance value no 
bigger than $\phi$ if the graph actually contains a cut with conductance 
$O(\phi^2)$~\cite{Cheeger69_bound,Chung:1997}.
This bound comes from a discrete version of Cheeger's inequality, which was
originally proved in a continuous setting for compact Riemannian manifolds; 
and it is parameterized in terms of a structural parameter of the input, but 
it is independent of the number $n$ of nodes in the graph.
\item
Finally, note that the worst-case quadratic approximation factor is 
\emph{not} an artifact of the analysis---it is obtained for spectral methods 
on graphs with ``long stringy'' pieces~\cite{guatterymiller98}, basically 
since spectral methods confuse ``long paths'' with ``deep cuts''---and that 
it is a very ``local'' property, in that it is a consequence of the 
connections with diffusion and thus it is seen in locally-biased versions of 
the spectral 
method~\cite{Spielman:2004,andersen06local,Chung07_heatkernelPNAS,MOV09_TRv3}.
\end{itemize}

Flow-based algorithms compute an approximation to 
Problem~(\ref{eqn:conductance_graph}) by 
solving an all-pairs multicommodity flow problem.
Several things are worth noting.
\begin{itemize}
\item
First, this multicommodity flow problem is a relaxation of 
Problem~(\ref{eqn:conductance_graph}), as can be seen by replacing the 
$x\in\{-1,1\}^{n}$ constraint (which provides a particular semi-metric) in 
the corresponding integer program with a general semi-metric constraint.
\item
Second, this procedure effectively embeds the data into an $\ell_1$ metric 
space, \emph{i.e.}, a real vector space $\mathbb{R}^{n}$, where distances are 
measured with the $\ell_1$ norm.
\item
Third, one can prove that the resulting partition is within an $O(\log n)$ 
factor of optimal, in the sense that the cut returned by the algorithm has 
conductance no bigger than $O(\log n)$, where $n$ is the number of nodes in 
the graph, times the conductance value of the optimal conductance set in the 
graph~\cite{LLR95_JRNL,Leighton:1999,HLW06_expanders}.
This bound comes from Bourgain's result which states that any $n$-point 
metric space can be embedded into Euclidean space with only logarithmic 
distortion, a result which clearly depends on the number $n$ of nodes in the 
graph but which is independent of any structural parameters of the~graph.
\item
Finally, note that the worst-case $O(\log n)$ approximation factor is 
\emph{not} an artifact of the analysis---it is obtained for flow-based 
methods on constant-degree expander 
graphs~\cite{LLR95_JRNL,Leighton:1999,HLW06_expanders}---and that it is a 
very ``global'' property, in that it is a consequence of the fact that for 
constant-degree expanders the average distance between all pairs of nodes 
is~$O(\log n)$.
\end{itemize}

Thus, spectral methods and flow-based methods are complementary in that 
they relax the combinatorial problem of optimizing conductance in very 
different ways;%
\footnote{For readers familiar with recent algorithms based on semidefinite 
programming~\cite{ARV_CACM08}, note that these methods may be viewed as 
combining spectral and flow in a particular way that, in addition to 
providing improved worst-case guarantees, also has strong connections with 
boosting~\cite{hast-tibs-fried}, a statistical method which in many cases 
is known to avoid over-fitting.  The connections with what I am discussing 
in this article remain to be explored.}
they succeed and fail for complementary input (\emph{e.g.}, flow-based 
methods do not confuse ``long paths'' with ``deep cuts,'' and spectral 
methods do not have problems with constant-degree expanders); and
they come with quality-of-approximation guarantees that are structurally 
very different.%
\footnote{These differences 
highlight a rather egregious 
theory-practice disconnect (that parallels the algorithmic-statistical 
disconnect).  In my experience, if you ask nearly anyone within theoretical 
computer science what is a good algorithm 
for partitioning a graph, 
they would say flow-based 
methods---after all flow-based methods run in low-degree polynomial time, 
they achieve $O(\log n)$ worst-case approximation guarantees, 
etc.---although they would note that spectral methods are better for 
expanders, basically since the quadratic of a constant is a constant.  On 
the other hand, nearly everyone outside of computer science would say 
spectral methods do pretty well for the data in which they are interested, 
and they would wonder why anyone would be interested in partitioning a 
graph 
without any 
good~partitions.}
For these and other reasons, spectral and flow-based approximation 
algorithms for the intractable graph partitioning problem provide a good 
``hydrogen atom'' for understanding more generally the disconnect between 
the algorithmic and statistical perspectives on data.  

Providing a precise statement of how spectral and flow-based approximation 
algorithms implicitly compute regularized solutions to the intractable 
graph partitioning problem (in a manner, \emph{e.g.}, analogous to how 
truncated diffusion-based procedures for approximating the leading 
nontrivial eigenvector of a graph Laplacian exactly solve a regularized 
version of the problem) has \emph{not}, to my knowledge, been accomplished.
Nevertheless, this theoretical evidence---\emph{i.e.}, that spectral and 
flow-based methods are effectively ``filtering'' the input data through very 
different metric and geometric places%
\footnote{That is, whereas traditional regularization takes place by solving 
a problem with an \emph{explicitly-imposed geometry}, where an explicit norm 
constraint is added to ensure that the resulting solution is ``small,'' one 
can view the steps of an approximation algorithm as providing an 
\emph{implicitly-imposed geometry}.
The details of how and where that implicitly-imposed geometry is ``nice'' 
will determine the running time and quality-of-approximation guarantees, as 
well as what input data are particularly challenging or well-suited for the 
approximation algorithm.}%
---suggests that this phenomenon exists.

To observe this phenomenon empirically, one should work with a class of data 
that highlights the peculiar features of spectral and flow-based methods, 
\emph{e.g.}, that has properties similar to graphs that ``saturate'' 
spectral's and flow's worst-case approximation guarantees.
Empirical evidence~\cite{LLDM08_communities_CONF,LLM10_communities_CONF} 
clearly demonstrates that large social and information networks have these 
properties---they are strongly expander-like when viewed at large size 
scales; their sparsity and noise properties are such that they have 
structures analogous to stringy pieces that are cut off or regularized away 
by spectral methods; and they often have structural regions that at least 
locally are meaningfully low-dimensional.
Thus, this class of data provides a good ``hydrogen atom'' for understanding 
more generally the regularization properties implicit in graph approximation 
algorithms.

\begin{figure}
   \begin{center}
      \subfigure[Objective function value]{
         \includegraphics[width=0.30\linewidth]{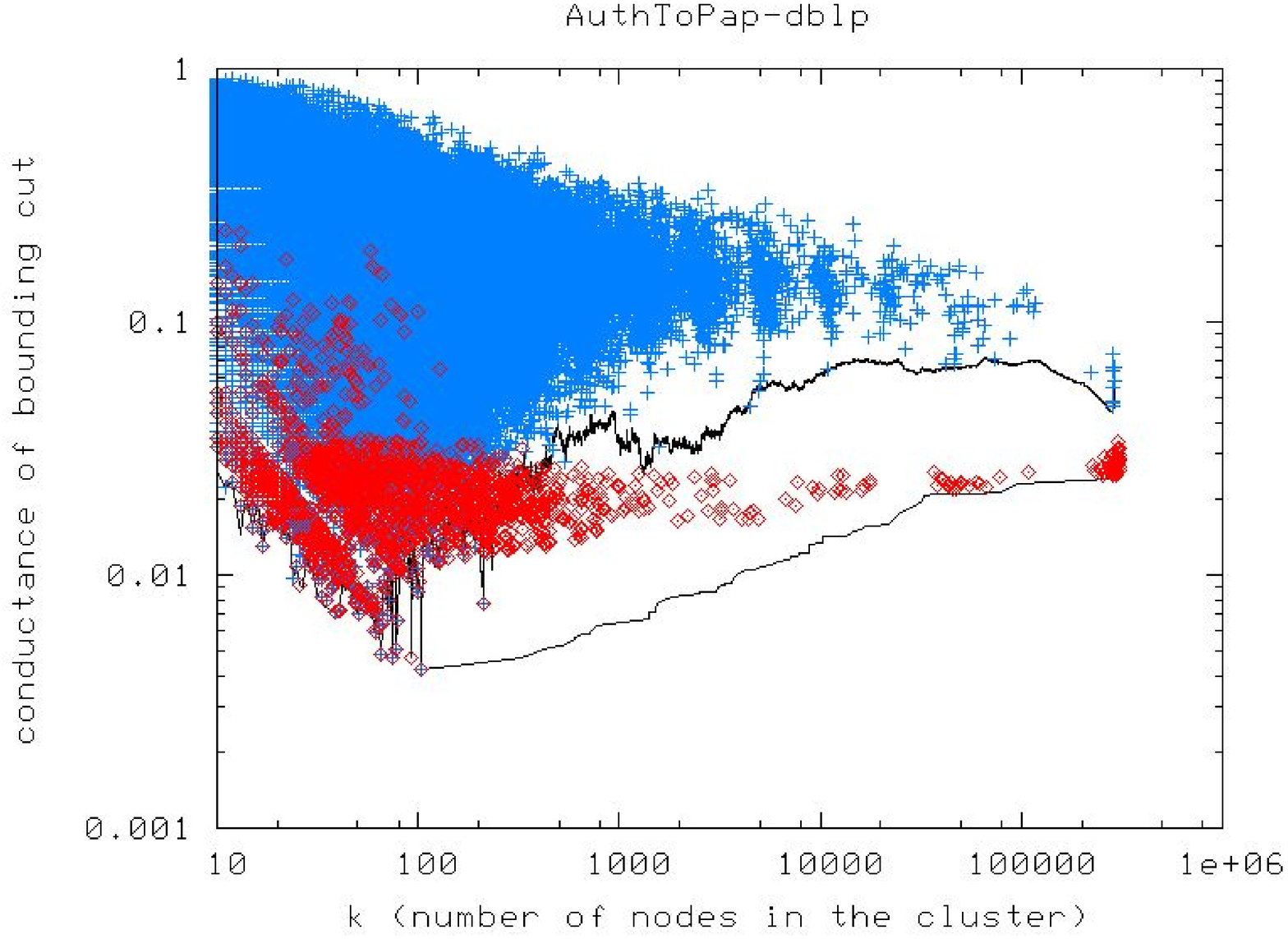}
         \label{compactness-vs-cuts-fig:obj}
      } %%% \qquad %%% &
      \subfigure[One ``niceness'' measure]{
         \includegraphics[width=0.30\linewidth]{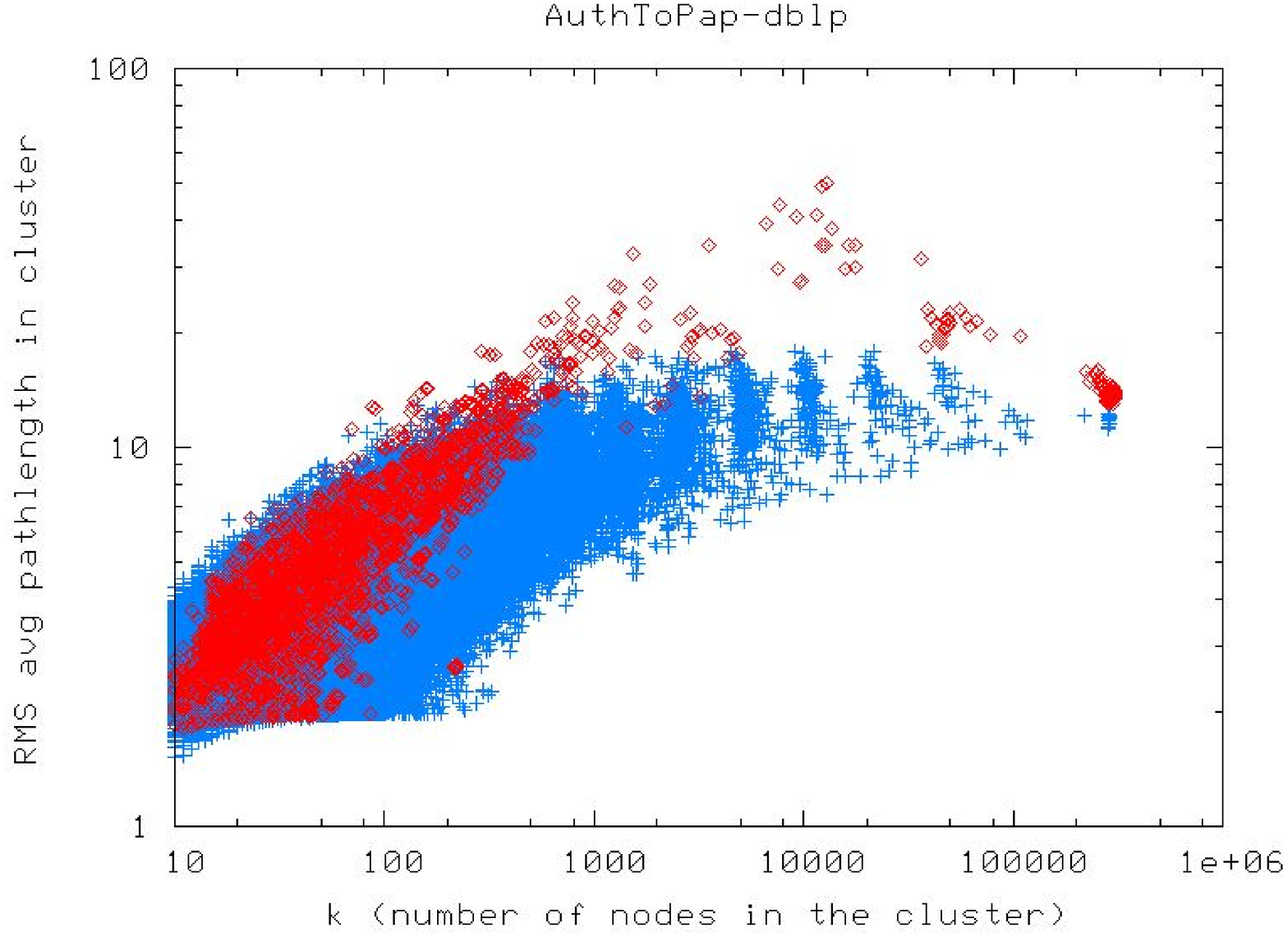}
         \label{compactness-vs-cuts-fig:nice1}
      } %%% \qquad %%% &
      \subfigure[Another ``niceness'' measure]{
         \includegraphics[width=0.30\linewidth]{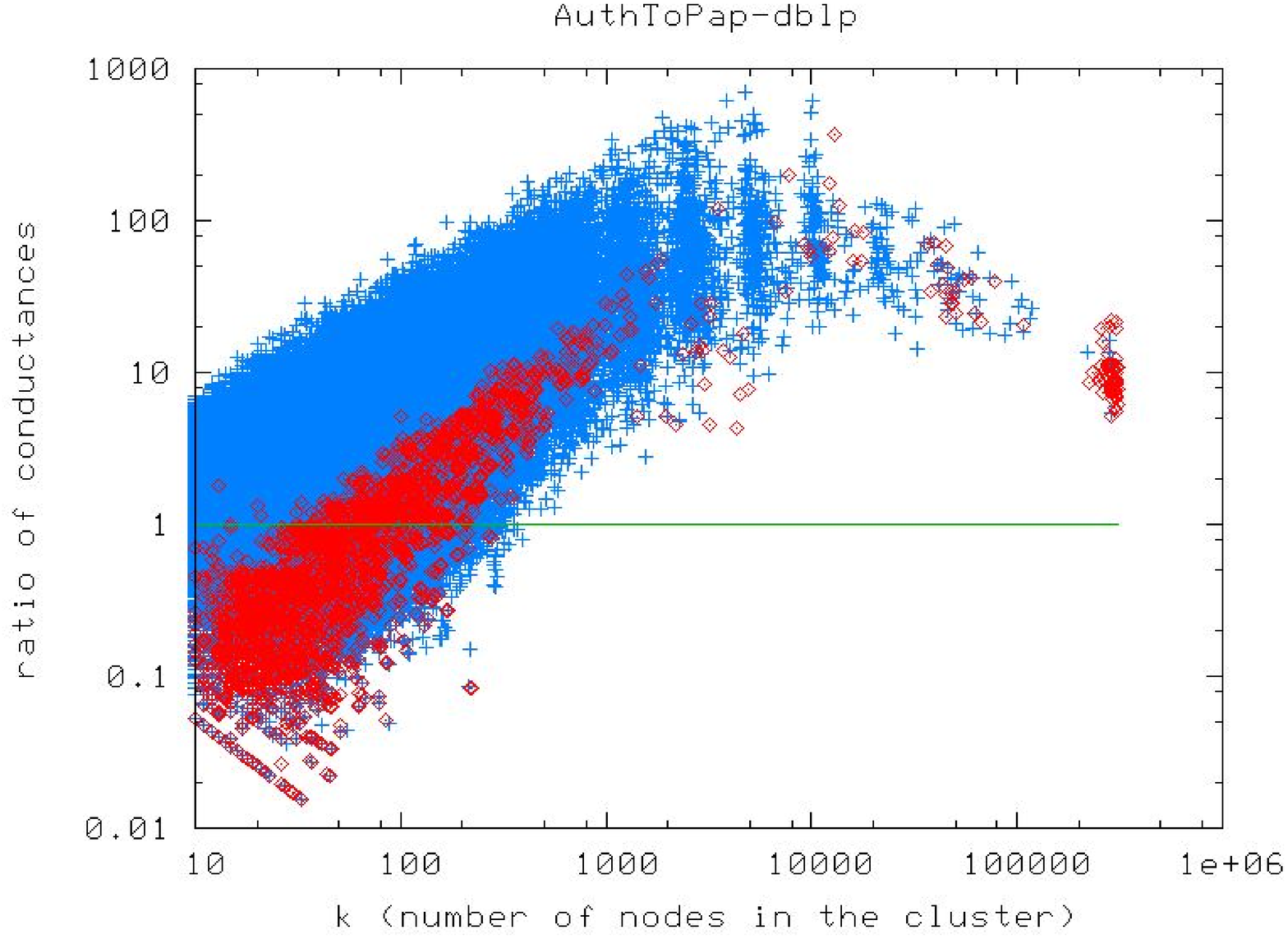}
         \label{compactness-vs-cuts-fig:nice2}
      } 
   \end{center}
\caption{
Scatter plot (on log-log scales) of size-resolved conductance 
(in Fig.~\ref{compactness-vs-cuts-fig:obj}) 
and two ``niceness'' measures 
(Fig.~\ref{compactness-vs-cuts-fig:nice1} shows average shortest path length 
and Fig.~\ref{compactness-vs-cuts-fig:nice2} shows the ratio of external 
conductance to the internal conductance)
for clusters found in the \textsc{AtP-DBLP} (\textsc{AuthToPap-dblp}) 
network with a spectral algorithm (blue) and a flow-based algorithm (red).
See~\cite{LLDM08_communities_CONF,LLM10_communities_CONF} for details.
For all plots, lower values of the Y-axis are ``better.''
In this and other examples, the flow-based algorithm (red, Metis+MQI) 
generally yields clusters with better conductance scores, while the spectral 
algorithm (blue, LocalSpectral) generally yields clusters that are~nicer.
}
\label{compactness-vs-cuts-fig}
\end{figure}

In light of this, let's say that we are interested in finding reasonably 
good clusters of size $10^3$ or $10^4$ nodes in a large social or information 
network.
(See~\cite{algstat10_CHAPTER} for why this might be interesting.) 
In that case, Figure~\ref{compactness-vs-cuts-fig} presents very typical 
results.
Figure~\ref{compactness-vs-cuts-fig:obj} presents a scatter plot of the 
size-resolved conductance of clusters found with a flow-based approximation 
algorithm (in red) and a spectral-based approximation algorithm (in blue).%
\footnote{Ignore the ``size-resolved'' aspect of these plots, since by 
assumption we are interested in clusters of roughly $10^3$ or $10^4$ nodes 
(but~\cite{LLDM08_communities_CONF,LLM10_communities_CONF} provides details 
on this); and don't worry about the details of the flow-based and spectral-based
procedures, except to say that there is a nontrivial theory-practice gap 
(again,~\cite{LLDM08_communities_CONF,LLM10_communities_CONF} provides 
details).}
In this plot, lower values on the Y-axis correspond to better values of the 
objective function; and thus the flow-based procedure is 
unambiguously
better than 
the spectral procedure at finding good-conductance clusters.
On the other hand, how useful these clusters are for downstream applications
is also of interest. 
Since we are not explicitly performing any regularization, we do not have 
any explicit ``niceness'' function, but we can examine empirical niceness
properties of the clusters found by the two approximation procedures.
Figures~\ref{compactness-vs-cuts-fig:nice1} 
and~\ref{compactness-vs-cuts-fig:nice2} presents these results for two 
different niceness measures.
Here, lower values on the Y-axis correspond to ``nicer'' clusters, and again
we are interested in clusters with lower Y-axis values.
Thus, in many cases,
the spectral procedure is clearly better than the flow-based procedure at 
finding nice clusters with reasonably good conductance values.

Formalizations aside, this empirical tradeoff between solution quality and 
solution niceness is basically the defining feature of statistical 
regularization---except that we are observing it here as a function of two 
different approximation algorithms for the same intractable combinatorial 
objective function.
That is, although we have not explicitly put any regularization term 
anywhere, the fact that these two different approximation algorithms 
essentially filter the data through different metric and geometric spaces 
leaves easily-observed empirical artifacts on the output of those 
approximation algorithms.%
\footnote{For other data---in particular, constant-degree expanders---the 
situation should be reversed.  That is, theory clearly predicts that 
locally-biased flow-based algorithms~\cite{andersen08soda} will have 
better niceness properties than locally-biased spectral-based 
algorithms~\cite{andersen06local}.  Observing this empirically on real 
data is difficult since data that are sufficiently unstructured to be 
expanders, in the sense of having no good partitions, tend to have very 
substantial degree heterogeneity.}
One possible response to these empirical results is is to say that 
conductance is not the ``right'' objective function and that we should 
come up with some other objective to formalize our intuition;%
\footnote{Conductance probably is the combinatorial quantity that most 
closely captures the intuitive bi-criterial notion of what it means for a 
set of nodes to be a good ``community,'' but it is still very far from perfect 
on many real data.}
but of course that other objective function will likely be intractable, and
thus we will have to approximate it with a different spectral-based or
flow-based (or some other) procedure, in which case the same implicit 
regularization issues will
arise~\cite{LLDM08_communities_CONF,LLM10_communities_CONF}.

\subsection{Computing locally-biased graph partitions}
\label{sxn:local}

In many applications, one would like to identify locally-biased graph 
partitions, \emph{i.e.}, clusters in a data graph that are ``near'' a 
prespecified set of nodes. 
For example, in nearly every reasonably large social or information 
network, there do not exist large good-conductance clusters, but there are 
often smaller clusters that are meaningful to the domain 
scientist~\cite{andersen06seed,LLDM08_communities_CONF,LLM10_communities_CONF}; 
in other cases, one might have domain knowledge about certain nodes, and one 
might want to use that to find locally-biased clusters in a semi-supervised 
manner~\cite{MOV09_TRv3};
while in other cases, one might want to perform algorithmic primitives such 
as solving linear equations in time that is nearly linear in the size of the 
graph~\cite{Spielman:2004,andersen06local,Chung07_heatkernelPNAS}.

One general approach to problems of this sort is to modify the usual 
objective function  and then show that the solution to the modified problem
inherits some or all of the nice properties of the original objective.
For example, 
a natural way to formalize the idea of a locally-biased version of the 
leading nontrivial eigenvector of $\mathcal{L}$ that can then be used in a 
locally-biased version of the graph partitioning problem is to modify 
Problem~(\ref{eqn:mo-unreg-vp}) with a locality constraint as follows.
\begin{equation}
\begin{aligned}
  & \underset{x}{\text{minimize}}
  & & x^T\mathcal{L}x \\ 
  & \text{subject to}
  & & x^Tx = 1, \\
  & & & x^T D^{1/2} 1 = 0  \\
  & & & (x^T D^{1/2} s)^2 \geq \kappa  ,
\end{aligned}
\label{eqn:mov-vp}
\end{equation}
where $s$ is a vector representing the ``seed set,'' and where $\kappa$ is a
locality parameter.
This \emph{locally-biased} version of the usual spectral graph partitioning 
problem was introduced in~\cite{MOV09_TRv3}, where it was shown that solution 
inherits many of the nice properties of the solution to the usual global 
spectral partitioning problem.
In particular, the exact solution can be found relatively-quickly by running 
a so-called Personalized PageRank computation;
if one performs a sweep cut on this solution vector in order to obtain a 
locally-biased partition, then one obtains Cheeger-like 
quality-of-approximation guarantees on the resulting cluster; and
if the seed set consists of a single node, then this is a relaxation of the 
following \emph{locally-biased graph partitioning problem}: 
given as input a graph $G=(V,E)$, an input node $u$, and a positive integer
$k$, find a set of nodes $S \subseteq V$ achieving
\begin{equation}
\phi(u,k,G) = \min_{S \subseteq V: u\in S, \mathrm{vol}(S) \le k} \phi(S) , 
\end{equation}
\emph{i.e.}, 
find the best conductance set of nodes of volume no greater 
than $k$ that contains the 
input 
node~$v$~\cite{MOV09_TRv3}. 
This ``optimization-based approach'' has the advantage that it is 
explicitly solving a well-defined objective function, and as such it is 
useful in many small-scale to medium-scale applications~\cite{MOV09_TRv3}.
But this approach has the disadvantage, at least for Web-scale graphs, that 
the computation of the locally-biased eigenvector ``touches'' all of the 
nodes in the graph---and this is very expensive, especially when one wants 
to find small clusters.  

An alternative more ``operational approach'' is to do the following: run 
some sort of procedure, the steps of which are similar to the steps of an 
algorithm that would solve the problem exactly; and then either use the 
output of that procedure in a downstream application in a manner similar to 
how the exact answer would have been used, or prove a theorem about that 
output that is similar to what can be proved for the exact answer.
As an example of this approach,
\cite{Spielman:2004,andersen06local,Chung07_heatkernelPNAS} take as input 
some seed nodes and a locality parameter and then run a diffusion-based 
procedure to return as output a ``good'' cluster that is ``nearby'' the 
seed nodes.
In each of these cases, the procedure is similar to the usual procedure,% 
\footnote{Namely, the three diffusion-based procedures that were described 
in Section~\ref{sxn:eigenvector}:
\cite{Spielman:2004} performs truncated random walks; 
\cite{andersen06local} approximates Personalized PageRank vectors; 
and~\cite{Chung07_heatkernelPNAS} runs a modified heat kernel procedure.}
except that at 
each step of the algorithm various ``small'' quantities are truncated to 
zero (or simply maintained at zero), thereby minimizing the number of nodes 
that need to be touched at each step of the algorithm.
For example,~\cite{Spielman:2004} sets to zero very small probabilities, 
and \cite{andersen06local} uses the so-called 
\emph{push algorithm}~\cite{JW03,Vig11_TR} to concentrate computational 
effort on that part of the vector where most of the nonnegligible changes 
will take~place.

The outputs of these \emph{strongly local spectral methods} obtain
Cheeger-like quality-of-approximation guarantees, and by design these 
procedures are extremely fast---the running time depends on the size of the
output and is independent even of the number of nodes in the graph.
Thus, an advantage of this approach is that it opens up the possibility of 
performing more sophisticated eigenvector-based analytics on Web-scale data 
matrices; and these
methods have already proven crucial in characterizing the clustering and 
community structure of social and information networks with up to millions of 
nodes~\cite{andersen06seed,LLDM08_communities_CONF,LLM10_communities_CONF}. 
At present, though, this approach has 
the disadvantage that it is very difficult to use: the exact statement of the 
theoretical results is extremely complicated, thereby limiting its
interpretability; 
it is extremely difficult to characterize and interpret for downstream 
applications what actually is being computed by these procedures, 
\emph{i.e.}, it is not clear what optimization problem these approximation 
algorithms are solving exactly;
and counterintuitive things like a seed node not being part of ``its own 
cluster'' can easily happen.
At root, the reason for these difficulties is that the truncation and 
zeroing-out steps implicitly regularize---but they are done based on 
computational considerations, and it is not known what are the implicit 
statistical side-effects of these design decisions.

The precise relationship between these two approaches has not, to my 
knowledge, been characterized.
Informally, though, the truncating-to-zero provides a ``bias'' that is 
analogous to the early-stopping of iterative methods, such as those 
described in Section~\ref{sxn:eigenvector}, and that has strong structural 
similarities with thresholding and truncation methods, as commonly used in 
$\ell_1$-regularization methods and
optimization more generally~\cite{FHT00}.
For example, the update step of the push algorithm, as used 
in~\cite{andersen06local}, is a form of stochastic gradient 
descent~\cite{GM-unpub}, a method particularly well-suited for large-scale 
environments due to its connections with regularization and 
boosting~\cite{bottou-2010}; and the algorithm terminates after a small 
number of iterations when a truncated residual vector equals 
zero~\cite{GM-unpub}, in a manner similar to other truncated gradient 
methods~\cite{LLZ09}.

Perhaps more immediately-relevant to database theory and practice as well as
to implementing these ideas in large-scale statistical data analysis 
applications is simply to note that this operational and interactive 
approach to database algorithms is \emph{already} being adopted in practice.
For example, in addition to empirical work that uses these methods to 
characterize the clustering and community structure of large 
networks~\cite{andersen06seed,LLDM08_communities_CONF,LLM10_communities_CONF}, 
the body of work that uses diffusion-based primitives in database environments 
includes an algorithm to estimate PageRank on graph streams~\cite{DGP08}, 
the approximation of PageRank on large-scale dynamically-evolving social 
networks~\cite{BCG10}, and a MapReduce algorithm for the approximation of 
Personalized PageRank vectors of all the nodes in a graph~\cite{BCX11}.

\section{Discussion and conclusion}
\label{sxn:conc}

Before concluding, I would like to share a few more general thoughts on 
approximation algorithm theory, in light of the above discussion.
As a precursor, I should point out the obvious fact that the modern theory of 
NP-completeness is an extremely useful theory.
It is a theory, and so it is an imperfect guide to practice; but it is a 
useful theory in the sense that it provides a qualitative notion of fast 
computation, a robust guide as to when algorithms will or will not perform 
well, etc.% 
\footnote{For readers familiar with Linear Programming and issues associated 
with the simplex algorithm versus the ellipsoid algorithm, it is probably 
worth viewing this example as the ``exception that proves the rule.''}
The theory achieved this by considering computation 
\emph{per se}, as a one-step process that divorced the computation from the 
input and the output except insofar as the computation depended on 
relatively-simple complexity measures like the size of the input.
Thus, the success is due to the empirical fact that many natural problems of 
interest are solvable in low-degree polynomial time, that the tractability 
status of many of the ``hardest'' problems in NP is in some sense 
equivalent, and that neither of these facts depends on the input data or 
the posedness of the problem.

I think it is also fair to say that, at least in a very wide range of MMDS
applications, the modern theory of approximation algorithms is nowhere near 
as analogously useful.  
The bounds the theory provides are often very weak; the theory often doesn't 
provide constants which are of interest in practice; the dependence of the 
bounds on various parameters is often not even qualitatively right; and in 
general it doesn't provide analogously qualitative insight as to when 
approximation algorithms will and will not be useful in practice for 
realistic noisy data.
One can speculate on the reasons---technically, the combinatorial gadgets 
used to establish approximability and nonapproximability results might not 
be sufficiently robust to the noise properties of the input data; 
many embedding methods, and thus their associated bounds, tend to emphasize 
the properties of ``far apart'' data points, while in most data applications 
``nearby'' information is more reliable and more useful for downstream 
analysis; 
the geometry associated with matrices and spectral graph theory is much more 
structured than the geometry associated with general metric spaces; 
structural parameters like conductance and the isoperimetric constant are
robust and meaningful and not brittle combinatorial constructions that 
encode pathologies; and   %% in general
ignoring posedness questions and viewing the analysis of approximate 
computation as a one-step process might simply be too coarse.

The approach I have described involves going ``beyond worst-case analysis'' 
to addressing questions that lie at the heart of the disconnect between what 
I have called the algorithmic perspective and the statistical perspective on 
large-scale data analysis.
At the heart of this disconnect is the concept of regularization,
a notion that is almost entirely absent from 
computer science, but which is central to nearly every application domain 
that applies algorithms to noisy data. 
Both
theoretical and empirical evidence demonstrates that approximate 
computation, in and of itself, can implicitly lead to statistical 
regularization, in the sense that approximate computation---either 
approximation algorithms in theoretical computer science or heuristic 
design decisions that practitioners must make in order to implement their 
algorithms in real systems---often implicitly leads to some sort of 
regularization.
This suggests treating statistical modeling questions and computational 
considerations on a more equal footing, rather than viewing either one as 
very much secondary to the~other.

The benefit of this perspective for database theory and the theory and 
practice of large-scale data analysis is that one can hope to achieve 
bicriteria of having algorithms that are scalable to very large-scale data 
sets and that also have well-understood inferential or predictive properties.
Of course, this is not a panacea---some problems are simply hard; some data 
are simply too noisy; and running an approximation algorithm may implicitly 
be making assumptions that are manifestly violated by the data.
All that being said, 
%%%
understanding and exploiting in a more principled manner the statistical 
properties that are implicit in scalable worst-case 
algorithms should be of interest in many very practical MMDS applications.

%\bibliographystyle{plain}
%\bibliography{communities,mwmbib_jrnl,mwmbib_proc,mwmbib_book,mwmbib_misc,mwmbib_drft,communities,geom-struct}

\end{document}